\documentstyle[twoside,fleqn,epsfig,espcrc2]{article}

\newcommand{\AmS}{{\protect\the\textfont2
   A\kern-.1667em\lower.5ex\hbox{M}\kern-.125emS}}

\title{On the scalar nonet lowest in mass}

\author{Peter Minkowski\address{Institute for Theoretical Physics,
     University of Bern,  CH-3012 Bern, Switzerland}
         \thanks{Work supported in part by Schweizerischer Nationalfonds}
         and
         Wolfgang Ochs\address{Max Planck Institut f\"ur Physik, Werner
      Heisenberg Institut,
            D-80805 Munich, Germany}}

\begin{document}

\begin{abstract}
The hypothesis that there exists a nonet of scalars mainly composed
of a valence quark-antiquark pair and mixed according to near 
singlet-octet separation : $f_{0} (980)$ singlet, 
\ $a_{0}^{+,0,-} (980) \ , \ K^{* \ +,0}_{0} (1430) \ , 
\ \overline{K}^{* \ -,0}_{0} (1430) \ f_{0} (1500)$ 
octet, is put to further tests from the three body decays 
$D^{\pm} , D_{s}^{\pm} \ \rightarrow \ PS^{\pm} \ \pi^{+} \pi^{-}$
with $PS^{\pm} \ = \ \pi^{\pm} \ , \ K^{\pm}$.
The analysis of decay phases supports the singlet nature of $f_{0} (980)$.
\end{abstract}

\maketitle

\section{Spectroscopy of $q \overline{q}$ p-waves}

\begin{figure}[htb]
\vspace*{-1.cm}
\begin{center}
\mbox{\epsfig{file=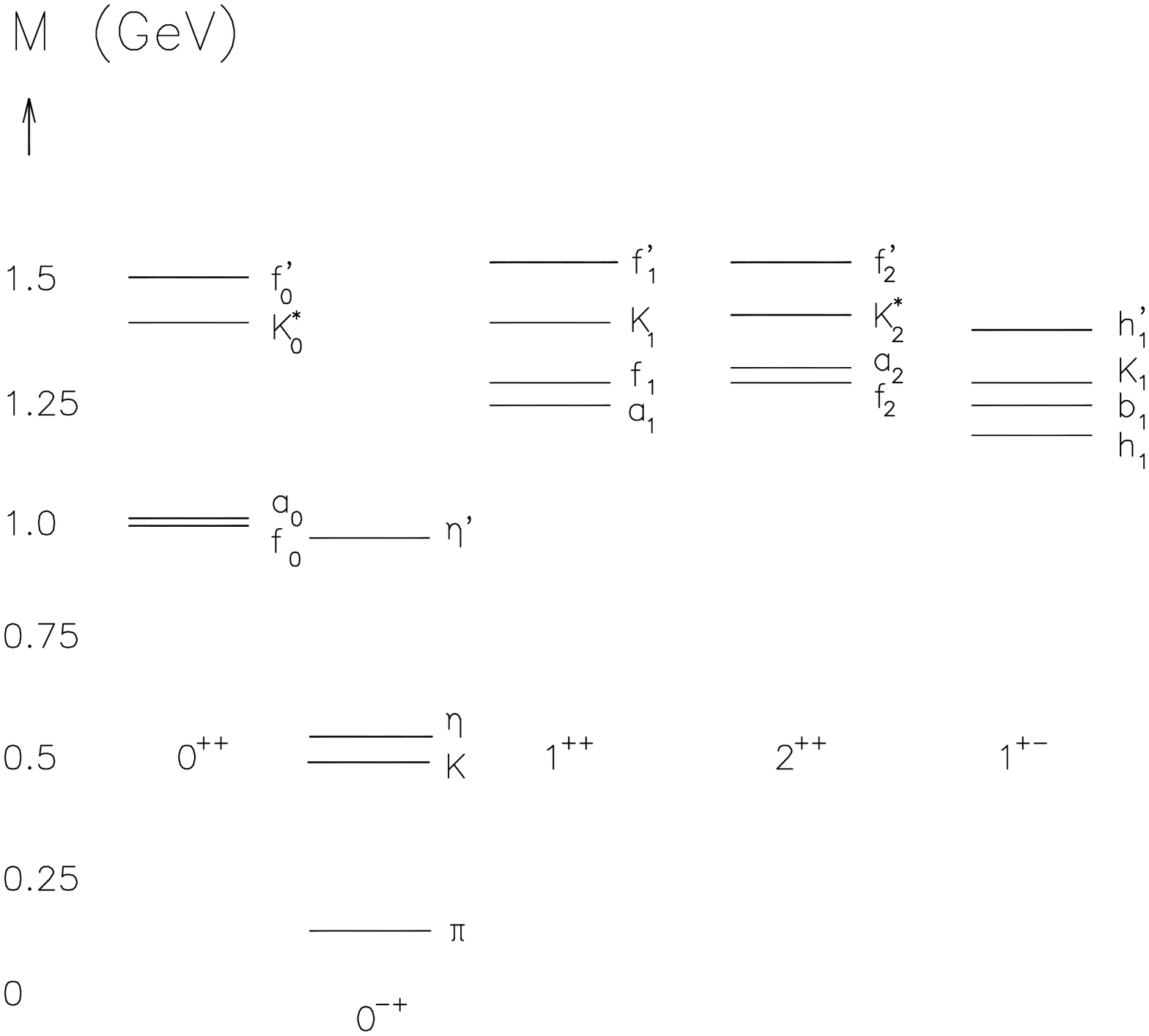,%
width=6.5cm,angle=90}}
\end{center}
\vspace*{-1.0cm}
\caption{$J^{PC_{n}} = 0^{++},0^{-+},1^{++},2^{++}, 1^{+-}$
$q \overline{q}$ nonets.}
\label{fig1}
\end{figure}

Following the hypothesis given in the abstract we select the following
4 p-wave nonets : 
$J^{PC_{n}} \ = \ 0^{++} \ , \ 1^{++} \ , \ 2^{++} \ , \ 1^{+-}$. 
They are displayed together with the pseudoscalar nonet in figure \ref{fig1}.

As a consequence of the adopted selection criteria we 
exclude the following candidate states from the PDG listings \cite{PDG}:
$f_{0} (400-1200)$, $f_{0} (1370)$, $f_{1} (1410)$ and 
resonances with a mass exceeding $f_{2}^{'} (1525)$ in particular
$h_{1} (1595)$ and $f_{0} (1710)$, see also \cite{PMWO}.
\vspace*{-0.0cm}

We pose the following questions\footnote{For a rather different
description of low lying scalar mesons we refer e.g. to ref. \cite{CloTo}.}:

\begin{description}
\item - Is the flavor mixing pattern of the $0^{-+}$ and
$0^{++}$ nonets similar , i.e. near singlet - octet? 
The two nonets would be parity doublets due to chiral symmetry,
which is spontaneously and explicitely broken.      

\item - Does the scalar gluonic meson distort through
large mixing effects the scalar $q \overline{q}$ nonet 
beyond (spectroscopic) recognition?

In our analysis in \cite{PMWO} we started from the hypothesis,
that the answer to this question is no. Inspecting the
spectra of the p-wave nonets in Fig. 1 we observe no strong distortion
in the scalar sector indeed.

\item - Can the last question be resolved through direct
observation of the scalar gluonic meson as a conventionally narrow
resonance and how reliable are the mass estimates from
purely gluonic lattice-QCD \cite{latQCD}

\begin{displaymath}
m_{gb} \ (0^{++}) \ = \ 1600 \ \mbox{MeV} 
\ \pm 10 \% \hspace*{0.3cm} \mbox{?}
\end{displaymath}

\item - Can QCD sum rules including 
local gluonic operators \cite{Nari} shed light on the above mass estimate?
The latter two questions are adressed in the accompanying paper
\cite{WO}.
\end{description}

{\bf The Gell-Mann - Okubo square mass formula} 

According to our analysis 
$f_{0} \ \rightarrow \ f_{<}$ (980) represents the SU3 singlet, whereas

$a_{0}$ (984.7) , $K_{0}$ (1412) , $f_{0} \ \rightarrow \ f_{>}$ (1507)

\noindent
form the associated octet.

\noindent
The Gell-Mann - Okubo (first order) mass square
relation then yields (in $\mbox{GeV}^{2}$ units)

\begin{equation}
\label{eq:2}
\begin{array}{l}
m^{2} (f_{>}) = 
\vspace*{0.2cm} \\
\hspace*{0.6cm} 
 m^{2} (a_{0}) + \frac{4}{3} 
 \left (  m^{2} (K_{0}) -  m^{2} (a_{0}) \right )
\vspace*{0.2cm} \\
2.271 \ = \ 0.970 \ + \ 1.365 \ = \ 2.335
\end{array}
\end{equation}

\noindent
the deviation amounts to $0.064/2.271 \ = \ 2.8 \ \%$
\footnote{Eq. \ref{eq:2} is numerically a
refinement with respect to ref. \cite{PMWO}.}

\noindent
There is no sign - yet - of any major distortion.

\noindent
The degeneracy in mass of $f_{<}$ and $a_{0}$, while
not offending any basic principles, indicates further
dynamic simplicity to be explained.

\section{Further evidence for (near) octet-singlet flavor 
phase structure} 

The aim is to study and eventually confirm
the nonstrange versus strange $q \overline{q}$ flavor
structure of the two singlet-octet assigned isoscalar scalars :

\begin{displaymath}
\begin{array}{lll ll}
f_{<} \ (980) & = & \cos \ \vartheta
\ \left | \ {\bf 0} \ \right \rangle
& - & \sin \ \vartheta 
\ \left | \ {\bf 8} \ \right \rangle
\vspace*{0.2cm} \\
f_{>} \ (1500) & = & \sin \ \vartheta
\ \left | \ {\bf 0} \ \right \rangle
& + & \cos \ \vartheta 
\ \left | \ {\bf 8} \ \right \rangle
\end{array}
\end{displaymath}

\noindent
The following phase convention shall be chosen in
the flavor basis

\begin{equation}
\label{eq:3}
\begin{array}{l}
f \ = \ \sum_{q} \ c_{q} 
\ \left | \ {\bf q \overline{q}} \ \right \rangle
\hspace*{0.1cm} , \hspace*{0.1cm} 
q \ = \ u,d,s
\vspace*{0.2cm} \\
c_{u}  =  c_{d}  =  \frac{1}{\sqrt{2}} \ c_{ns}
\hspace*{0.0cm} ; \hspace*{0.0cm} 
c_{ns}  =  \sin \varphi , c_{s}  =  \cos \varphi
\vspace*{0.2cm} \\
 \left | \ {\bf 0} \ \right \rangle
 =  \frac{1}{\sqrt{3}} \left (  1 , 1 , 1  \right )
\hspace*{0.1cm} ; \hspace*{0.1cm} 
 \left | \ {\bf 8} \ \right \rangle
 =  \frac{1}{\sqrt{6}} \left (  1 , 1 , -2  \right )
\vspace*{0.2cm} \\
 \left | \ {\bf ns} \ \right \rangle
\ = \ \frac{1}{\sqrt{2}} \left (  1 , 1 , 0  \right )
\hspace*{0.1cm} ; \hspace*{0.1cm} 
 \left | \ {\bf s} \ \right \rangle
 =  \left (  0 , 0 , 1  \right )
\end{array}
\end{equation}

\noindent
and in the ns-s basis we have

\begin{equation}
\label{eq:4}
\begin{array}{l}
 \left | \ {\bf 0} \ \right \rangle
\ = \ \sin \ \varphi_{*} 
 \left | \ {\bf ns} \ \right \rangle
\ + \ \cos \ \varphi_{*} 
\ \left | \ {\bf s} \ \right \rangle
\vspace*{0.2cm} \\
 \left | \ {\bf 8} \ \right \rangle
\ = \ \cos \ \varphi_{*} 
\ \left | \ {\bf ns} \ \right \rangle
\ - \ \sin \ \varphi_{*}
\ \left | \ {\bf s} \ \right \rangle
\vspace*{0.2cm} \\
\varphi_{*} \ = \ \mbox{arccot} \ \frac{1}{\sqrt{2}} \ = \ 54.74^{o}
\end{array}
\end{equation}

\noindent
so in the flavor basis we have

\begin{equation}
\label{eq:5}
\begin{array}{l}
f_{<} \ (980) = \sin \varphi
\ \left |  {\bf ns}  \right \rangle
 +  \cos \varphi 
\ \left | {\bf s} \right \rangle
\vspace*{0.2cm} \\
f_{>} \ (1500)  =  \cos \varphi
\ \left | {\bf ns} \right \rangle
 -  \sin \varphi 
\ \left | {\bf s}  \right \rangle
\vspace*{0.2cm} \\
\vartheta \ = \ \varphi_{*} \ - \ \varphi 
\end{array}
\end{equation}

\noindent
As conjectured range of the
singlet-octet angle $\vartheta$ we consider

\begin{equation}
\label{eq:6}
\begin{array}{l}
0 \ \leq \ \vartheta \ \leq \ \vartheta_{m}
\ , \ \vartheta_{m}  =  \mbox{arcsin} \frac{1}{3} \ = \ 19.47^{o}
\end{array}
\end{equation}

\noindent
The corresponding range for $\varphi$ becomes

\begin{equation}
\label{eq:7}
\begin{array}{l}
35.26^{o} \ \leq \ \varphi \ \leq \ \varphi_{*}
\ = \ 54.74^{o}
\end{array}
\end{equation}

\noindent
Previously \cite{PMWO} we have analysed the decays involving $f_{0} (980)$ :
$J/\Psi \rightarrow \phi f_{0}, \omega f_{0}$, the
radiative decays $f_{0},a_{0} \rightarrow \gamma \gamma$
and $f_{0} \rightarrow K \overline{K}, \pi \pi$ and concluded
on a large flavor mixing similar to $\eta$-$\eta^{'}$ in the pseudoscalar
nonet. Now we extend our analysis to D decays.

\noindent
Our analysis of the mixing pattern shall focus
on the two ratios of strange to nonstrange components

\begin{equation}
\label{eq:8}
\begin{array}{l}
R_{<} \ = \ R \ ( f_{0} (980) ) \ = \ \mbox{cot} \ \varphi
\vspace*{0.2cm} \\
R_{>} \ = \ R \ ( f_{0} (1500) ) \ = \ - \ \mbox{tg} \ \varphi
\end{array}
\end{equation}

\noindent
If the premise of mainly singlet $f_{0}$ (980) is correct
we infer the ranges $R_{<} \ > \ 0$ and
$R_{>} \ = \ - \ 1 / R_{<} < 0$ . The relation
$R_{<} \ R_{>} \ = \ -1$ follows from orthogonality.

\noindent
The restricted ranges are

\begin{equation}
\label{eq:9}
\begin{array}{rll lr}
\frac{1}{\sqrt{2}} & \leq  & R_{<} & \leq & \sqrt{2}
\vspace*{0.2cm} \\
- \ \sqrt{2} & \leq & R_{>} & \leq & - \ \frac{1}{\sqrt{2}}
\end{array}
\end{equation}

\noindent
We determine $\varphi$ considering the three three body
decays of the charmed mesons $D$ and $D_{s}$ \cite{E791}:


\begin{equation}
\label{eq:10}
\begin{array}{l}
A) \ D^{+}_{s} \ \rightarrow \ f_{0} \ (980) \ \pi^{+} 
\vspace*{0.2cm} \\
B) \ D^{+} \ \rightarrow \ f_{0} \ (980) \ \pi^{+} 
\vspace*{0.2cm} \\
C) \ D^{+} \ \rightarrow \ K_{0}^{*} \ (1430) \ \pi^{+} 
\end{array}
\end{equation}


 \begin{figure}[htb]
\begin{center}
\mbox{\epsfig{angle=90,file=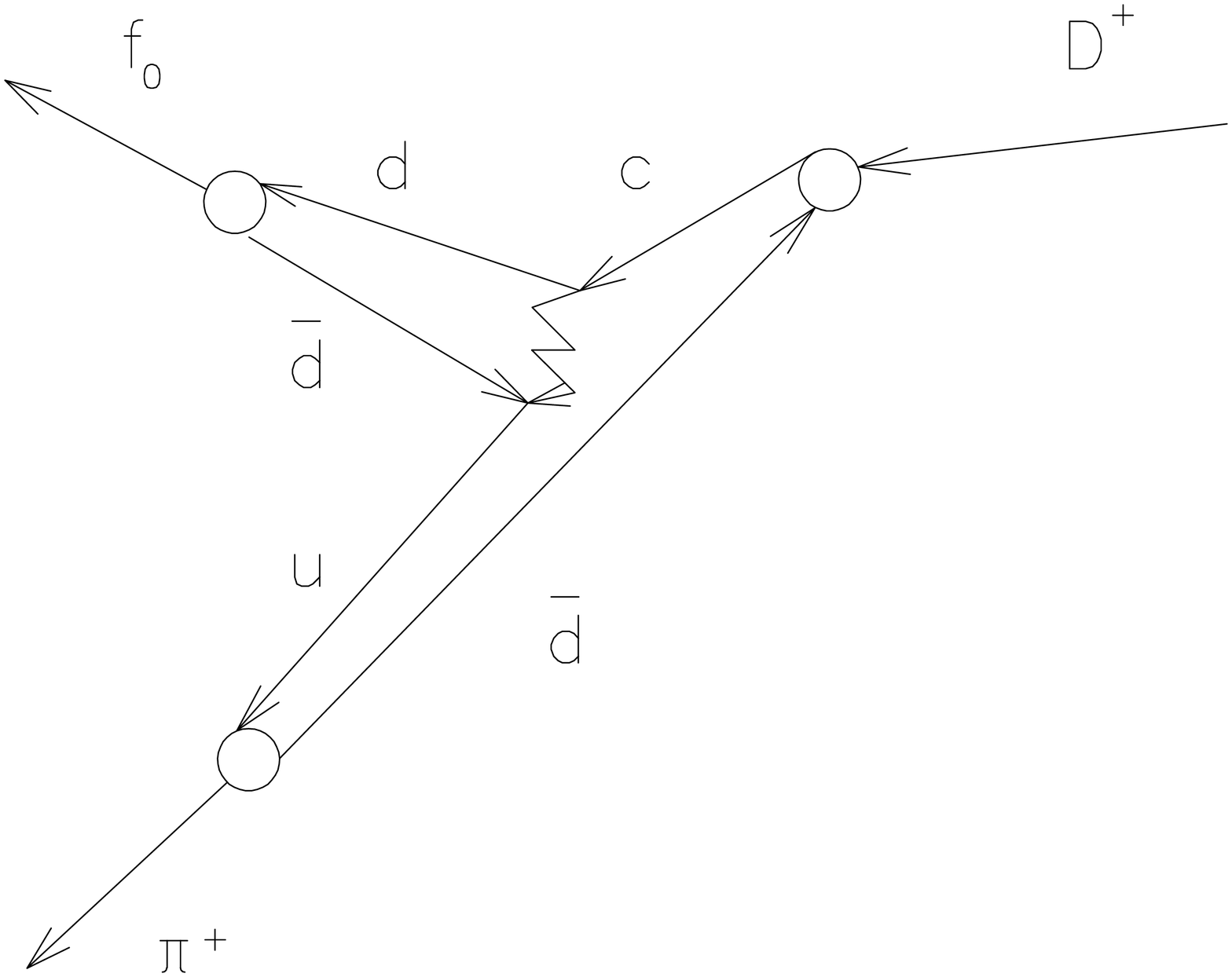,width=6.cm}}
\mbox{\epsfig{angle=90,file=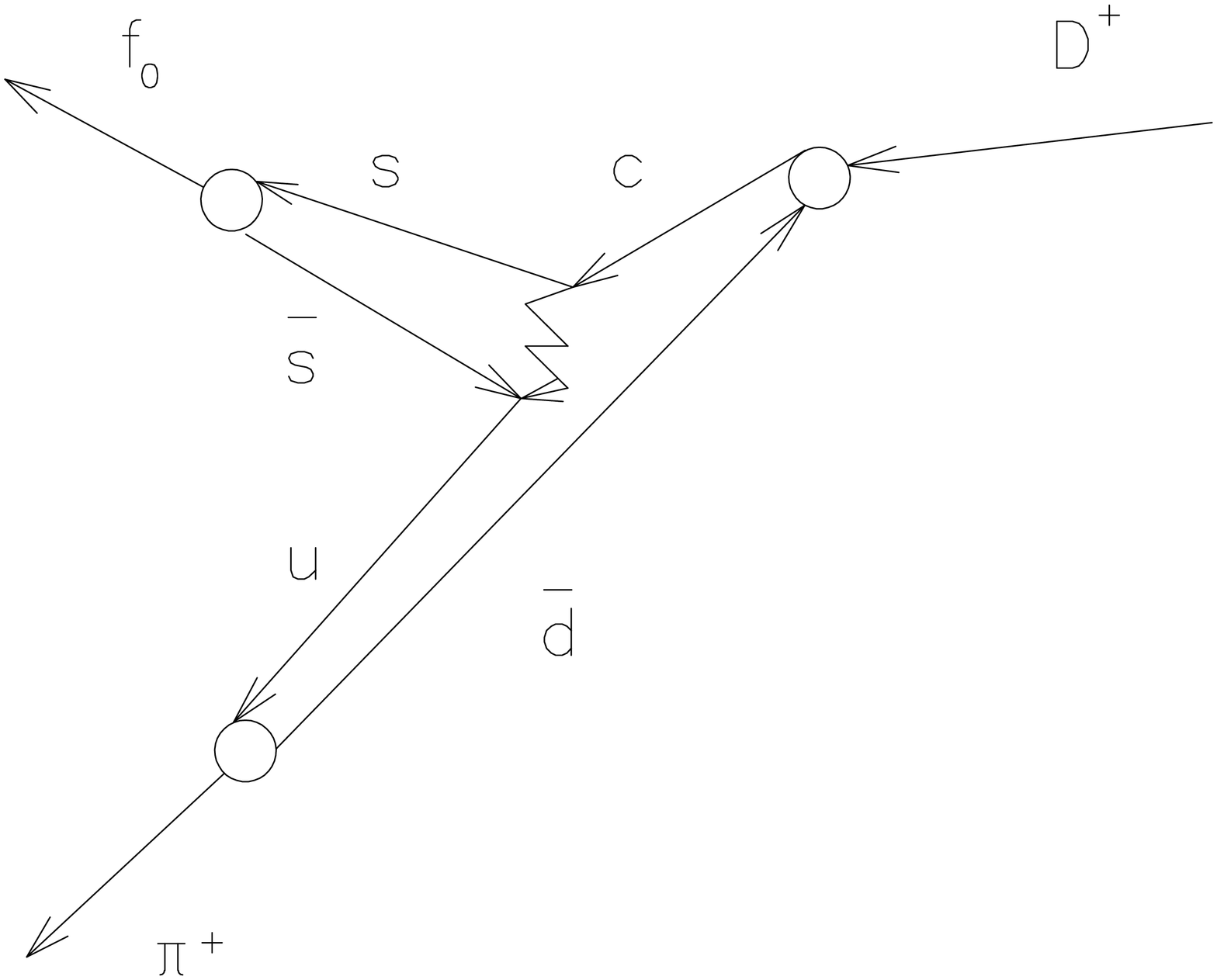,width=6.cm}}
\end{center}
\caption{The two color suppressed quark flavor flow diagrams ($\propto
\epsilon a$) in reaction B.}
\label{fig4}
 \end{figure}
\vspace*{0.0cm}




\noindent
We consider the color favored amplitudes ($\propto$ a)
which contribute to all processes in eq. \ref{eq:10},
and color suppressed amplitudes 
($\propto$ $\epsilon a$) obtained by 
$\overline{d}_{1} \ \leftrightarrow \ \overline{d}_{2}$ 
(see Fig. \ref{fig4}) and obtain


\begin{equation}
\label{eq:11}
\begin{array}{l}
A \ = \ \cos\varphi \ V_{ud}V_{cs}^* \ a 
\vspace*{0.2cm} \\
B \ = \ \frac{\sin\varphi}{\sqrt{2}}
           V_{ud}V_{cd}^* 
 (1+\epsilon)  a  +  \cos\varphi  V_{us}V_{cs}^*
 \epsilon \ a 
\vspace*{0.2cm} \\
C \ =  \ V_{ud}V_{cs}^*  \  (1+\epsilon) \ a 
  \label{damp}
\end{array}
\end{equation}

\noindent
We consider the two ratios of partial decay widths $A / B$
and $A / C$ :

\begin{equation}
\label{eq:12}
\begin{array}{l}
R_{A/B}  =  \frac{\Gamma(D_s^+\to f_0(980) \pi^+)} 
{\Gamma(D^+\to f_0(980) \pi^+)}
\vspace*{0.2cm} \\
R_{A/C}  =  \frac{\Gamma(D_s^+\to f_0(980) \pi^+)} 
{\Gamma(D^+\to K_0(1430) \pi^+)}
\end{array}
\end{equation}

\noindent
In the approximation $V_{ud}=V_{cs}=\cos\vartheta_c$,
$V_{us}=-V_{cd}=\sin\vartheta_c$ with Cabibbo angle $\vartheta_{c}$,
we find

\begin{equation}
\label{eq:13}
\begin{array}{l}
R_{A/B}  =  2 \frac{\Phi_1}{\Phi_2} \cot^2\vartheta_c\cot^2\varphi
     \frac{1}{|1-(\sqrt{2}\cot\varphi -1)\epsilon |^2}
\vspace*{0.2cm} \\
R_{A/C}  =   \frac{\Phi_1}{\Phi_3} \cos^2\varphi
     \ \frac{1}{|1+\epsilon |^2}
\label{r4r5}
\end{array}
\end{equation}

\noindent
where $\Phi_{1,2,3} = (p_{\pi^+})_{1,2,3}$ denote the phase space in
s-wave decays, proportional to the $\pi^+$ momentum in the decay resonance
rest frame.
Using the branching fractions established by the E791 Collaboration
and the PDG results we find numerically

\begin{equation}
\label{eq:14}
\begin{array}{l}
\cot^2\varphi  /  |  1-(\sqrt{2}\cot\varphi -1)\epsilon  |^2 
 = 
\vspace*{0.2cm} \\
\hspace*{1.0cm} = 1.26 \ (1.0 \pm 0.4 )
\vspace*{0.2cm} \\
\cos^2 \varphi \ / \ | \ 1+\epsilon  \ |^2  
 =  0.52 \ ( 1.0 \pm \ 0.3 )
\label{angle1}
\end{array}
\end{equation}

\noindent
Then we obtain as solutions two bands (a and b)
due to the quadratic nature of the relations in eq. \ref{eq:14}

\begin{equation}
\label{eq:15}
\begin{array}{l}
 \cot \varphi_{a}  =  1.11^{\ + 0.33}_{\ - 0.20}
\hspace*{0.1cm} , \hspace*{0.1cm} 
\varphi_{a} = \left . 42.14^{\ + 5.8}_{\ - 7.3} \right .^\circ
\vspace*{0.2cm} \\
 \epsilon_{a}  =  ( 2.85 \ \pm \ 5.35 ) \ 10^{-2} 
\vspace*{0.2cm} \\
 \tan \varphi_{b}  =  - 0.34^{\ + 0.46}_{\ - 0.37}
\hspace*{0.1cm} , \hspace*{0.1cm} 
\varphi_{b} = \left . 161.2^{\ + 25.7}_{\ - 16.5} \right .^\circ
\vspace*{0.2cm} \\
 \epsilon_{b}  =   0.31^{\ + 0.004}_{\ - 0.13} 
\label{phieps}
\end{array}
\end{equation}

The angle $\varphi$ is defined modulo $180^{\circ}$.
The two solutions in eq. \ref{eq:15} can be distinguished
through the sign of the quantity $R_{<}$ (or $R_{>}$) defined
in eq. \ref{eq:8}.

The phase (+ --) structure of $f{<}$ , $f_{>}$
is determined from interference with other resonances in 
$D^{+}$ and $D^{+}_{s}$ decays
into $3 \pi$ and $\pi K \overline{K}$ \cite{E791,E687}.
The amplitudes A , B , C in eq. \ref{eq:11} exhibit the  
+ -- phase structure shown in table 1.

\begin{table*} 
\caption{Decay phases (in degrees) for resonances in $D$ and $D_s$ decays
as measured by experiments E687 
and E791 . In each line an overall phase
 has been fixed arbitrarily. The states marked by ($*$) are not directly
evident in the plots.
Comparison with theoretical expectations for the mixing angle $\varphi$
and predictions for $f_0(980)$ near flavour
singlet and $f_0(1500)$ as octet partner
($0^\circ<\varphi<90^\circ$) with arbitrary angle $\alpha$,
standard choice is $\alpha=0$. }
\vspace*{0.3cm}

$
 \begin{array}{lc@{\hspace*{3.2cm}}c@{\hspace*{3.2cm}}c}
 \hline
 D\to 3\pi   &d \bar d \to  & &
  \\
             &  \rho(770)        &  f_0(980)         & f_2(1270)^*
  \\
\mbox{E687}  &  27\pm 14 \pm 11  & 197 \pm 28 \pm 24 & 207 \pm 17 \pm 4 \\
\mbox{E791}  & 0 (\mbox{fixed})  & 151.8 \pm 16.0      & 102.6 \pm 16.0  \\
\mbox{Theory}& -d\bar d          & d\bar d \sin(\varphi) & d\bar d \\
             & \alpha+180^\circ  &  \alpha           & \alpha \\
\hline
 D_s\to 3\pi  &  s \bar s \to & &   \\
             &  f_0(980)         & f_2(1270)^*         &f_0(1500)
  \\
\mbox{E687}  & 0(\mbox{fixed})   & 83\pm 16          & 210 \pm 10   \\
\mbox{E791}  & 0(\mbox{fixed})   & 133 \pm 13\pm 28  & 198 \pm 19 \pm 27 \\
\mbox{Theory}& s\bar s \cos(\varphi)  & \varepsilon s\bar s &
                   - s\bar s \sin(\varphi)\\
             & \alpha           &  \alpha          & \alpha+180^\circ\\
\hline
D_s\to \pi K\bar K \qquad  &  s \bar s \to & &   \\
             &  f_0(980)^*         & \phi(1020)  &
  \\
\mbox{E687}  &  159 \pm 22 \pm 16 & 178 \pm 20\pm 24 &    \\
\mbox{Theory}& s\bar s \cos(\varphi)  & s\bar s    &    \\
             &   \alpha           &   \alpha         &    \\
\hline
  \end{array} 
$
\label{tab:phases}
\end{table*}

\section{Conclusions and outlook}

It becomes clear from the results in table 1 and the assumed
form of the amplitudes in eq. \ref{eq:11}, that only the solution
in band a) in eq. \ref{eq:15} is compatible with the data.
This implies 
$\varphi = \left . 42.14^{\ + 5.8}_{\ - 7.3} \right .^\circ$
confirming the near singlet quark flavor mixing of $f_{0} (980)$,
the ideal singlet corresponds to
$\varphi_{*} = 54.74^\circ$.

The present analysis leaves the mixing with the 
scalar gluonic meson(s)
completely open. Here we refer to our present results
on glueballs in ref. \cite{WO}. Future work will hopefully
establish the full structure of the 
scalar nonet lowest in mass including the scalar glueball $gb (0^{++}$).

\section*{Acknowledgement}

It is a pleasure to thank all organizers and participants
of the QCD2002 conference, in particular Stephan Narison,
for enabling the meeting to evolve in a
lively and fruitful atmosphere.

\end{document}